\documentclass[a4paper, prb, twocolumn]{revtex4}

\usepackage{amssymb}
\usepackage{amsmath}
\usepackage{psfig}
\usepackage{color}
\usepackage{graphics, graphicx}
\usepackage{psfrag}

\begin{document}
\title{Microscopic Wave Functions of Spin Singlet and Nematic Mott States
\\ of Spin-One Bosons in High Dimensional Bipartite Lattices}

\author{Michiel Snoek}
\author{Fei Zhou \footnote{Permanent address:
Department of Physics and Astronomy,\\
University of British Columbia,
6224 Agriculture Road, Vancouver, B.C. V6T
1Z1,Canada}}
\affiliation{ITP, Utrecht University, Minnaert building, Leuvenlaan 4,
3584 CE Utrecht, The Netherlands}
\date{\today}

\begin{abstract}
We present microscopic wave functions of spin singlet Mott
insulating states and nematic Mott insulating states.
We also investigate quantum phase transitions between
the spin singlet Mott phase and the nematic Mott phase in both
large-$N$ limit and small-$N$ limit
($N$ being the number of particles per site) in high dimensional 
bipartite lattices.
In the mean field approximation employed in this article
we find that phase transitions are generally weakly first order.
\\
\\
PACS numbers: 03.75.Mn, 75.10.Jm, 75.45.+j
\end{abstract}

\maketitle

\section{Introduction}

The recent observation of correlated states of bosonic atoms in 
optical lattices has generated much interest.\cite{Greiner02, Greiner02b}
As known for a while, when bosons in lattices interact with each other 
repulsively, they can be localized and form a Mott insulating 
state instead of a condensate. \cite{Fisher89,Jaksch98}
This phenomenon has been observed in the optical lattice 
experiment. By varying laser intensities of optical lattices, 
Greiner et al. have successfully 
investigated Mott states of spinless bosons by
probing spin polarized cold atoms in optical 
lattices with a large potential depth. \cite{Greiner02, Greiner02b}

We are interested in  spin correlated Mott insulating 
states of spin-one bosons, especially spin-one bosons with 
antiferromagnetic interactions.
Some aspects of spin correlated Mott insulating states 
were investigated recently.
For an even number of particles per site,
both spin singlet Mott insulators and nematic Mott insulators  
were found in certain parameter regimes,  
while  for high dimensional lattices with an odd number of particles per 
site
only nematic insulating states
were proposed. \cite{Demler01}
In one-dimensional lattices, it was demonstrated that for an
odd number of particles per site, Mott states should be dimerized 
valence-bond-crystals,
which support interesting fractionalized quasi-excitations. \cite{Zhou02}
Effects of spin correlations on Mott insulator-superfluid
transitions have been studied and remain to be fully
understood. \cite{Tsuchiya02}

In this article, we analyze the microscopic structures of spin singlet
Mott insulating states (SSMI) and nematic Mott insulating states (NMI).
We study, quantitatively, quantum phase transitions
between these two phases in
high-dimensional bipartite
lattices. In the mean
field approximation, we demonstrate that for an even number of particles
per site, the transitions are weakly first order.
We should emphasize that results obtained in this paper are only valid in
high dimensions. In one dimensional lattices,
nematic order does not survive long wave length quantum fluctuations;
detailed discussions on low dimensional Mott states for both even and odd numbers
of particles per site are presented in
Ref. 6.

The organization is as follows. In section II, we present
the general setting for the study of spin order-disorder quantum
phase transitions. In section III, we present mean field results on
the quantum phase transitions in both small $N$ and large $N$ limits.
In section IV, we
discuss issues which are to be understood in the future.

\section{Algebra and setting}
\subsection{The microscopic Hamiltonian in the dilute limit}
The microscopic lattice Hamiltonian we employ to study spin correlated 
states of spin-one bosons is:
\begin{eqnarray}
\mathcal{H}_{\text{microscopic}} &=& - t \sum_{\langle k l \rangle} (\psi_{k,
m}^\dagger \psi_{l, m} + \text{h.c.}) \nonumber \\
&&+ \sum_{k,l}
\psi_{k, m}^\dagger \psi_{k, m} U^{\rho}(k , l) 
\psi_{l, m'}^\dagger \psi_{l, m'} \\
&&+  \sum_{k,l} \psi_{k,  m}^\dagger {S}_{m n}^\gamma 
\psi_{k n} U^S(k, l)
\psi_{l, m'}^\dagger {S}_{m' n'}^\gamma 
\psi_{l, n'} \nonumber
\end{eqnarray} 
Here $\psi_{k, m}^\dagger$ is the creation operator of a spin-one particle 
at site
$k$ with spin-index $m=0, \pm 1$. $\langle k l \rangle$ indicates that the 
sum should be
taken over
nearest neighbors and ${S}^\gamma$ ($\gamma=x,y,z$) are spin-one matrix
operators given as:
\begin{equation*}
 S^x \!=\! \frac{1}{\sqrt{2}} \!\! \left( \!\! \begin{array}{ccc} 0 &1 &0 
 \\ 1 & 0 & 1
\\ 0 & 1 & 0 \end{array} \!\! \right) 
S^y\!=\!\frac{1}{\sqrt{2}} \!\! \left( \!\! \begin{array}{ccc} 0 &-i &0 \\ 
i & 0 & -i
\\ 0 & i & 0 \end{array} \!\!\! \right) 
S^z\!=\! \left( \!\! \begin{array}{ccc} 1 &0 &0 \\ 0 & 0 & 0
\\ 0 & 0 & \!-1 \end{array} \!\!\! \right).
\end{equation*}
$U^{\rho}(k,l)$ and $U^{s}(k,l)$ are, respectively, spin-independent and 
spin-dependent interaction parameters between two bosons at site $k$ and
$l$.

In the dilute limit, which is defined as a limit where $\bar \rho a^3 \ll 
1$ ($a$ is the scattering length and $\bar \rho$ the average density), 
atoms scatter in $s$-wave channels. For two spin-one atoms, the scattering 
takes place in the total spin $S=0,2$ channels, with scattering lengths 
$a_{0,2}$. Interactions between atoms can be approximated as spin-dependent 
contact interactions. \cite{Ho98}
In the lattice model introduced here, calculations yield
\begin{equation}
U^\rho (k, l) = E_c \delta_{k l} \quad \text{and} \quad
U^S (k, l) = E_s \delta_{k l}. 
\end{equation}  
The parameters $E_c$ and $E_s$ are given by:
\begin{equation} \label{cees}
E_c = \frac{ 4 \pi \hbar^2 \bar \rho (2 a_2+ a_0) }{3 M N} \tilde c, \quad
E_s = \frac{ 4 \pi \hbar^2 \bar \rho (a_2 - a_0) }{3 M N} \tilde c,
\end{equation} 
where $N$ is the average number of atoms
per site, $M$ is the mass of atoms and $\tilde c$ is a constant.

\subsection{Algebras}
For the study of spin correlated states in lattices, it is rather 
convenient to introduce the following operators:
\begin{subequations}
\begin{eqnarray}
\psi_{k,x}^\dagger &=& \frac{1}{\sqrt{2}} ( \psi_{k, -1}^\dagger -
\psi_{k, 1}^\dagger ) \\
\psi_{k,y}^\dagger &=& \frac{i}{\sqrt{2}} ( \psi_{k, -1}^\dagger 
+ \psi_{k, 1}^\dagger ) \\
\psi_{k,z}^\dagger &=& \psi_{k,0}^\dagger   
\end{eqnarray}
\end{subequations} 
where $k$ again labels a lattice site. In this representation:
\begin{equation}
\psi_{k,  m}^\dagger {S}_{m n}^\alpha \psi_{k, n}
= \hat {\bf S}_k^\alpha \equiv - i \epsilon^{\alpha \beta \gamma} 
\psi_{k, \beta}^\dagger \psi_{k, \gamma} 
\end{equation}
$\alpha, \beta, \gamma \in \{x,y,z \}$.
The density operator can be expressed in a usual way:
\begin{equation}
\hat \rho_k \equiv \psi_{k, \gamma}^\dagger \psi_{k, \gamma}.
\end{equation}
Consequently the Hamiltonian is given as:
\begin{eqnarray} \label{Hlatt}
\mathcal{H}_{\text{latt}} \!\!=\! - \tilde t \sum_{\langle k l \rangle}
 ( \psi_{k, \alpha}^\dagger \psi_{l, \alpha} \!\! + \text{h.c.} ) 
 \!+\! 
 \sum_k \!
 E_c \hat \rho_k^2 \!+\! E_s \hat {\bf S}_k^2 \! - \! \hat \rho_k \mu
\end{eqnarray}
where we have introduced the chemical potential $\mu$; $\hat {\bf S}_k^2$ 
is the total spin operator $\hat {\bf S}_{k, \alpha} \hat {\bf S}_{k, 
\alpha}$.

$\psi_{k, \alpha}$ ($\alpha=x,y,z$) are bosonic operators obeying the 
following commutation relations:
\begin{equation}
\lbrack \psi_{k, \alpha}, \psi_{l, \beta} \rbrack = 
\lbrack \psi_{k, \alpha}^\dagger, \psi_{l, \beta}^\dagger \rbrack = 0, \;\;
\lbrack \psi_{k, \alpha}, \psi_{l, \beta}^\dagger \rbrack = \delta_{k l}
\delta_{\alpha \beta}. 
\end{equation}
Taking into account Eqs. 5,6,8, one can verify the following algebras:
\begin{subequations}
\begin{eqnarray}
\lbrack \hat {\bf S}_k^\alpha, \psi_{l, \beta} \rbrack &=& \delta_{k l} 
i \epsilon^{\alpha \beta \gamma} \psi_{k, \gamma} \\
\lbrack \hat {\bf S}_k^\alpha, \psi_{l, \beta}^\dagger \rbrack &=& 
\delta_{k l} 
i \epsilon^{\alpha \beta \gamma} \psi_{k, \gamma}^\dagger \\
\lbrack \hat {\bf S}_k^\alpha, \hat {\bf S}_l^\beta \rbrack &=& 
\delta_{k l} i \epsilon^{\alpha \beta \gamma} \hat {\bf S}_k^\gamma \\
\nonumber \\
\lbrack \hat \rho_k, \psi_{l, \alpha} \rbrack &=& -\delta_{ k l} \psi_{k,
\alpha} \\
\lbrack \hat \rho_k, \psi_{l, \alpha}^\dagger \rbrack 
&=& \delta_{ k l} \psi_{k, \alpha}^\dagger \\
\nonumber \\
\lbrack \hat {\bf S}_k^\alpha, \hat \rho_l \rbrack &=&0 
\end{eqnarray}
\end{subequations}

Of particular interest is the singlet creation operator 
\begin{equation}
\frac{1}{\sqrt{6}}\psi_{k,
\alpha}^\dagger \psi_{k, \alpha}^\dagger=\frac{1}{\sqrt{6}}(\psi_{k,
0}^\dagger \psi_{k,0}^\dagger-2 \psi_{k,1}^\dagger \psi_{k,-1}^\dagger). 
\end{equation}
We find the following properties for this operator:
\begin{subequations}
\begin{eqnarray}
\lbrack \hat {\bf S}_k^\alpha, \psi_{l, \alpha}^\dagger 
\psi_{l,\alpha}^\dagger \rbrack &=&
\lbrack \hat {\bf S}_k^\alpha, \psi_{l, \alpha} \psi_{l,\alpha} \rbrack = 
0;
\\
\lbrack \psi_{k, \alpha} \psi_{k, \alpha}, \psi_{l, \beta}^\dagger \rbrack
&=&
2 \delta_{k l} \psi_{k, \beta}; \\
\lbrack \psi_{k, \alpha} \psi_{k, \alpha}, \psi_{l, \beta}^\dagger 
\psi_{l, \beta}^\dagger \rbrack &=& \delta_{k l} (4 \hat \rho_k + 6). 
\label{singsing}
\end{eqnarray} 
\end{subequations}

\subsection{The on-site dynamics}
The total spin operator can be expressed as:
\begin{equation}
\hat {\bf S}_k^2 = \hat \rho_k (\hat \rho_k+1) - \psi_{k, \alpha}^\dagger
\psi_{k, \alpha}^\dagger \psi_{k, \beta} \psi_{k, \beta}.
\end{equation}
So, eigenstates of the total spin operator
have to be eigenstates of the "singlet counting operator"
$\psi_{k, \alpha}^\dagger \psi_{k, \alpha}^\dagger \psi_{k, \beta}
\psi_{k, \beta}$.

Defining the state $\Psi_{k,0}^n$ such that:
\begin{equation}
\hat \rho_k  \Psi_{k,0}^n= n \Psi_{k,0}^n, \quad 
\psi_{k, \alpha}^\dagger \psi_{k, \alpha}^\dagger \psi_{k, \beta} \psi_{k, 
\beta} 
\Psi_{k,0}^n = 0,
\end{equation}
we find that wave functions of these eigenstates are:
\begin{equation}
\Psi_{k,m}^n = C (\psi_{k, \alpha}^\dagger \psi_{k, \alpha}^\dagger)^m
\Psi_{k,0}^{n-2m}
\end{equation}
where $C$ is a normalization constant. From Eq. \ref{singsing} it 
follows that:
\begin{equation}
\psi_{k, \alpha}^\dagger \psi_{k, \alpha}^\dagger \psi_{k, \beta} 
\psi_{k, \beta} 
\Psi_{k,m}^n = (4m(n-m)+2m) \Psi_{k,m}^n
\end{equation}
Using that $\hat \rho_k \Psi_{k,m}^n = n \Psi_{k,m}^n$ we derive:
\begin{equation}
\hat {\bf S}_k^2 \Psi_{k,m}^n = (n-2m)(n-2m+1) \Psi_{k,m}^n.
\end{equation}
So $S_k=n-2m$. Now if $n$ is even, $S_k$ is also even and when $n$ is odd, 
$S_k$ is
odd too. For an even number of particles per site $N$ the
states labeled by $S_k=0,2,4,\ldots, N$ are present, whereas for an
odd number of
particles per site $S_k=1,3,5,\ldots, N$ are allowed. This reflects the 
basic property of the many
body wave function of spin-one bosons, which has to be symmetric under the 
interchange of
two particles.

Solutions for spin correlated condensates
with finite numbers of particles were previously obtained \cite{Law98};
in the thermodynamical limit, these states evolve into
polar condensates. \cite{Ho98,Ohmi98, Castin00}
Also there, two-body scatterings were shown to lead to either
"antiferromagnetic" or "ferromagnetic" spin correlations in condensates.
Spin correlated condensates have been investigated in experiments.
\cite{Stamper-Kurn98,Stenger98}

\subsection{The effective Hamiltonian for Mott states}
In the limit when $\tilde t \ll E_c$, atoms are localized and only virtual 
exchange processes
are allowed. An effective Hamiltonian in this limit can be
derived in a second order perturbative calculation of the Hamiltonian in 
Eq.\ref{Hlatt}: 
\begin{equation} \label{Heff}
\mathcal{H}_\text{\tiny Mott} =\sum_k \frac{\hat {\bf S}_k^2}{2I} - 
\tilde J_\text{ex} \sum_{
\langle k l \rangle} \left(  \psi_{k, \alpha}^\dagger \psi_{l, \alpha} 
\psi_{l, \beta}^\dagger \psi_{k, \beta} + \text{h.c.} \right).
\end{equation}  
Here $\tilde J_\text{ex} = \frac{\tilde t^2}{2 E_c}$. 
In deriving Eq. \ref{Heff}, we have taken into account that $E_s \ll E_c$.

To facilitate discussions, we introduce the following operator:
\begin{equation}
\hat Q_{k, \alpha \beta} = \psi_{k, \alpha}^\dagger \psi_{k, \beta} -
\frac{1}{3} \delta_{\alpha \beta} \psi_{k, \gamma}^\dagger \psi_{k, 
\gamma}, 
\end{equation}
whose expectation value 
\begin{equation}
\tilde Q = \frac{ \langle \hat Q_{\alpha \beta} \rangle}{\langle \hat 
Q_{\alpha
\beta} \rangle_{\text{ref}}}
\end{equation}
is the nematic order parameter. The reference state $\psi_{\text{ref}} =
\prod_k \frac{ ({\bf n}_{\alpha} \psi_{k,\alpha}^\dagger)^N}{\sqrt{N!}} | 0 \rangle$
is a maximally ordered state. Choosing ${\bf n}= {\bf e}_z$, we obtain:
\begin{equation}
\langle \hat Q_{\alpha \beta} \rangle_{\text{ref}} = N \left(
\begin{array}{ccc} -\frac{1}{3} & 0 & 0 \\ 0 & -\frac{1}{3} & 0 \\ 0 & 0 &
\frac{2}{3} \end{array} \right) .
\end{equation}
$\tilde Q$ varies in a range of $\lbrack -\frac{1}{2}, 1 \rbrack$.

In terms of the operator $\hat Q_{\alpha \beta}$, the effective Mott
Hamiltonian can be rewritten as (up to an energy shift):
\begin{equation}
\mathcal{H}_\text{eff} = E_s \sum_k \text{Tr} \lbrack \hat Q_k \hat Q_k -
\hat
Q_k \hat Q_k^\dagger \rbrack - \tilde J_\text{ex} \sum_{\langle k l \rangle}
\text{Tr}
\lbrack \hat Q_k \hat Q_l \rbrack.
\end{equation}
Finally we define
\begin{equation}
\tilde \eta = \frac{z \tilde J_\text{ex}}{E_s}
\end{equation}
as a dimensionless parameter, which can be varied continuously; $z$ is the
coordination number of lattice.

\subsection{The range of the physical parameters}
From Eq. \ref{cees} it is clear that $E_s$ and $E_c$ depend on the
density, number of atoms, the mass of atoms and scattering lengths.
However, their ratio depends only on the scattering lengths. According to
current estimates \cite{Burke98, Ciobanu02}, for sodium atoms this ratio
is given as $\frac{E_s}{E_c}
\approx 9 \cdot 10^{-2}$. In this paper, we are interested in the limit
$E_s \ll E_c$.

The parameter $\tilde t$ can be varied independently by changing the depth
of the optical lattice.
A wide range is experimentally
accessible; one can vary from the regime where $\tilde t \gg E_c$ to a
regime where $\tilde t \ll E_s$. We limit ourselves to Mott states
($\tilde t \ll E_c$), where all bosons are localized, but the ratio $\tilde \eta$
can have arbitrary values.

\section{Phase transitions between SSMI's and NMI's}
\subsection{Two particles per site}

In the case of two particles per site ,
the on-site Hilbert space is six-dimensional, including a spin singlet
state

\begin{equation}
|S=0,S_z=0 \rangle  = \frac{\psi_\eta^\dagger
\psi_\eta^\dagger}{\sqrt{6}} | 0 \rangle,
\end{equation}
and five spin $S=2$-states
\begin{equation}
|Q_{\eta \xi} \rangle = \frac{\sqrt{3}}{2}
Q_{\eta \xi} \psi_{\eta}^\dagger
\psi_{\xi}^\dagger | 0 \rangle
\end{equation}
where $Q_{\eta \xi}$ is a symmetric and traceless tensor with five
independent elements.
All states in the Hilbert space are symmetric
under the interchange of bosons;
as expected, the states $|Q_{\eta\xi}\rangle$ are orthogonal to $|S=0,
S_z=0 \rangle$. It is convenient to choose the following representation of
$Q_{\eta\xi}$:

\begin{equation}
Q_{\eta \xi} ({\bf n}) = {\bf n}_{\eta} {\bf n}_{\xi} - \frac{1}{3}
\delta_{\eta \xi},
\end{equation}
with the director $\bf n$ as a unit vector living on $S^2$.
States defined by the director ${\bf n}$ form an over-complete set in the
subspace spanned by five $S=2$ states.

When the hopping is zero, one notices that the Hamiltonian in
Eq.\ref{Heff} commutes with $\hat {\bf S}^2_k$;
the ground state wave function is
\begin{equation}
|\Psi \rangle = \prod_k |S=0, S_z=0
 \rangle_k.
\end{equation}
On the other hand,
when $E_s$ goes to zero,
the Hamiltonian commutes with $\text{Tr} \lbrack \hat{Q}_{k,\alpha\beta}
\hat{Q}_{l,\beta\alpha} \rbrack $ and the
ground state wave function can be confirmed as:
\begin{equation}
|\Psi \rangle = \prod_k
\sqrt{\frac{2}{3}} |Q({\bf n}) \rangle_k + \frac{1}{\sqrt{3}} |S=0, S_z=0
\rangle_k.
\end{equation}
for any choice of the director $\bf n$.

To study spin nematic or spin singlet Mott
states at an arbitrary $\tilde \eta$,
we introduce a trial wave function
which is a linear superposition of singlet
states and symmetry breaking states:

\begin{equation} \label{theta}
|\Psi\rangle_\theta = \prod_k
\cos \theta | S=0, S_z=0 \rangle_k + \sin \theta |
Q_{\eta \xi} ({\bf n} ) \rangle_k .
\end{equation}
Here $\theta$ is a variable to be determined by the variational method.

A straightforward calculation leads to the following results:

\begin{eqnarray}
E ({\theta}) &=& \langle \Psi| \mathcal H | \Psi
\rangle_\theta  \\ &=&
6 E_s \sin^2 \theta - z \tilde J_{\text{ex}} \frac{2}{3} (
(2 \sqrt{2} \cos \theta \sin \theta + \sin^2 \theta )^2 \nonumber \\
\tilde Q &=&
( \sqrt{2} \cos \theta \sin \theta + \frac{\sin^2 \theta}{2} )
\end{eqnarray}
In terms of $\tilde Q$, the energy can be
expressed as:
\begin{equation*}
E  = 6 E_s \left( \frac{4}{9} + \frac{2}{9} \tilde Q -\frac{4}{9}
\sqrt{ - 2 \tilde Q^2 + \tilde Q + 1} \right) - \frac{8}{3} z \tilde J_{\text{ex}}
\tilde Q^2,
\end{equation*}
which for $\tilde Q \ll 1$ can be expanded as:
\begin{equation}
\left( 3 E_s - \frac{8}{3} z \tilde J_{\text{ex}} \right) \tilde Q^2
 - \frac{3}{2} E_s \tilde Q^3  + \frac{39}{16} E_s \tilde Q^4 + \ldots
\end{equation}
The cubic term leads to a first order phase transition in the mean field
approximation, which is similar to the situation in classical nematic
liquid crystals. \cite{deGennes}

\begin{figure}
\begin{center}
\psfrag{tilde_Q}{$\tilde Q$}
\psfrag{EoverEs}{$E/E_s$}
\includegraphics[scale=.8]{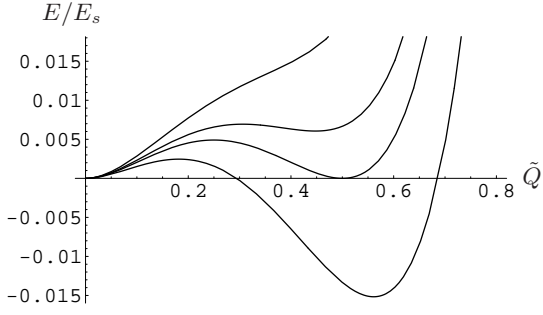}
\caption{Energy (measured in units of $E_s$) versus
$\tilde{Q}$ for various $\tilde \eta$ for $N=2$.  Curves from top to
bottom are for $\tilde \eta=0.97$, $0.99$, $1.0$, $1.02$.}
\label{energy}
\end{center}
\end{figure}

In FIG.\ref{energy} the $\tilde Q$-dependence of energy is plotted for
various $\tilde \eta$ in the vicinity of a quantum
critical point (mean field).
For $\tilde \eta < 0.985$, the energy has only one minimum at $\tilde{Q}=0$ and
correspondingly the ground state is a spin singlet Mott state.
When $0.985 < \tilde \eta < 1.0$, in addition to the global minimum at $\tilde
Q=0$, there appears a local minimum at $\tilde{Q} > 0$, which represents a
spin nematic metastable state.
When $\tilde \eta>1.0$ the solution with $\tilde Q > 0$ becomes a global minimum
and the solution at $\tilde Q = 0 $ is metastable; consequently the ground
state is a nematic Mott state.
For $\tilde \eta>\frac{9}{8}$, the solution at $\tilde Q=0$ becomes unstable ;
but an additional local minimum appears
at $\tilde Q<0$ which we interpret as a new metastable state (not shown in
FIG. \ref{energy}).

\begin{figure}
\begin{center}
\psfrag{tilde_Q}{$\tilde Q$}
\psfrag{tilde_eta}{$\tilde \eta$}
\includegraphics[scale=.9]{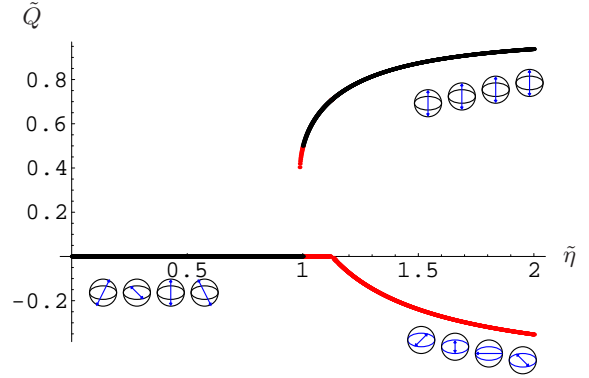}
\caption{(Color online) The nematic order parameter as a function
of $\tilde \eta$ for $N=2$. The phase transition takes place at $\tilde \eta=1$. Data
along
the black lines represent ground states; the red (light) lines are for metastable
states.
Spheres with double-headed arrows are introduced to represent ordering in
director $\bf n$ defined in Eq. 25 in different Mott states. In spin
singlet states, the director $\bf n$ is uncorrelated; in rod-like nematic
states, the director $\bf n$ is ordered and in disk-like states, the axis
of the easy plane of the director $\bf n$ is ordered.}
\label{qeta}
\end{center}
\end{figure}

The evolution of ground states as $\tilde \eta$ is varied, is summarized in
FIG. \ref{qeta}.
As is clearly visible, the phase transition is a  weakly first order one.
The jump in $\tilde Q$ at the phase-transition ($\tilde \eta=1.0$) is equal to
$\frac{1}{2}$.

It is worth emphasizing that
a positive $\tilde Q$ corresponds to a rod-like nematic
state;  for $\tilde Q=1$ the state is microscopically given by:
\begin{equation}
\frac{({\bf n}_{\alpha} \psi_{\alpha}^\dagger )^2}{\sqrt{2}} | 0 \rangle.
\end{equation}

A  solution with negative
$\tilde Q$ indicates a disk-like nematic state; the microscopic wave
function is
\begin{equation} \left(
\frac{1}{2} \psi_\eta^\dagger \psi_\eta^\dagger -
\frac{({\bf n}_{\alpha} \psi_{\alpha}^\dagger )^2}{2} \right) | 0 \rangle.
\end{equation}
at $\tilde Q=-\frac{1}{2}$.
For ${\bf n} = {\bf e}_z$ the wave functions in Eq.32,33 become $\frac{1}{\sqrt{2}}
\psi_z^\dagger \psi_z^\dagger | 0 \rangle$ and $\frac{1}{2} \left(
\psi_x^\dagger \psi_x^\dagger + \psi_y^\dagger \psi_y^\dagger \right)
|0\rangle$ respectively.

We have also tried a five-parameter variational approach, taking into 
account the full on-site Hilbert space. 
In a slightly different representation we write the trial wave function
as:
\begin{eqnarray} \label{psifull}
|\Psi \rangle &=& \prod_k 
\left( c_{xx} |xx \rangle_k + c_{yy} |yy \rangle_k + c_{zz} |zz \rangle_k 
\right.
\nonumber \\ & & \left. +
c_{xy} |xy \rangle_k + c_{xz} |xz \rangle_k + c_{yz} |yz \rangle_k \right)
\end{eqnarray}
Here $|\alpha \alpha \rangle_k = \frac{1}{\sqrt{2}} \psi_{k,\alpha}^\dagger
\psi_{k, \alpha}^\dagger |0\rangle$ (no summation) and $|\alpha \beta 
\rangle_k = \psi_{k,\alpha}^\dagger
\psi_{k,\beta}^\dagger| 0 \rangle$.
This results in the following expression for the energy:

\begin{eqnarray}
E &=& E_s \left[ 
4(c_{xx}^2+c_{yy}^2+c_{zz}^2-c_{xx}c_{yy}-c_{xx}c_{zz}-c_{yy}c_{zz}) 
\right.
\nonumber \\ & & \left. +
     6(c_{xy}^2+c_{xz}^2+c_{yz}^2) \right] \nonumber \\
&& - z \tilde J_{\text{ex}} \left[ 6(c_{xx}^4+c_{yy}^4+c_{zz}^4)
+4(c_{xy}^4+c_{xz}^4+c_{yz}^4)
\right. \nonumber \\ && 
+4(c_{xx}^2 c_{yy}^2+c_{xx}^2 c_{zz}^2+ c_{yy}^2 c_{zz}^2 )
 \nonumber \\
&& +12(c_{xx}^2 c_{xy}^2 + c_{xx}^2 c_{xz}^2 + c_{yy}^2 c_{xy}^2 + 
c_{yy}^2c_{yz}^2
\nonumber \\ && \quad \quad \quad
+c_{zz}^2 c_{xz}^2 + c_{zz}^2 c_{yz}^2)
\nonumber \\ &&
+ 8( c_{xy}^2 c_{xz}^2 + c_{xy}^2 c_{yz}^2 + c_{xz}^2 c_{yz}^2) \nonumber 
\\
&& 
+ 8 \sqrt{2} ( c_{xx} + c_{yy} + c_{zz}) c_{xy} c_{xz} c_{yz} 
\nonumber \\ &&
+ 4 (c_{xx}^2c_{yz}^2 + c_{yy}^2 c_{xz}^2 + c_{zz}^2 c_{xy}^2)
\nonumber \\ && \left.
+ 8 ( c_{xx} c_{yy} c_{xy}^2 + c_{xx} c_{zz} c_{xz}^2 + c_{yy} c_{zz} 
c_{xz}^2 )
\right].
\end{eqnarray} 

The conclusions are almost the same and summarized below: \\
i) For $\tilde \eta < 0.985 $. the only minimum is at
$c_{xx}=c_{yy}=c_{zz}=\frac{1}{\sqrt{3}}$, $c_{\alpha \beta}=0$ for 
$\alpha \neq \beta$. \\
ii) At $\tilde \eta=0.985$ additional local minima appear. \\
iii) At $\tilde \eta=1$ a first order  phase transition takes place. \\
iv) For $\frac{8}{9}>\tilde \eta>1$, the global minimum is at $\tilde Q > 0$, but 
the $\tilde Q=0$-solution remains to be a local minimum. \\
v) At $\tilde \eta=\frac{9}{8}$ the solution at 
$c_{xx}=c_{yy}=c_{zz}=\frac{1}{\sqrt{3}}$
becomes unstable. \\
vi) However, the disk-like $\tilde Q<0$-solution appears in this case as a 
saddle point.

\subsection{Large $N$ limit: An even number of particles per site}
For a large number of particles per site, it is convenient to introduce
the following
coherent state representation:
\begin{equation}
|{\bf n}, \chi \rangle = \frac{1}{\sqrt{2 \delta N}} \!\! \sum_{m=N-\delta
N}^{N+\delta N} \!\!\! \exp(-i m \chi) \frac{ \left( {\bf n}_\alpha
\psi_\alpha^\dagger \right)^m }{
\sqrt{2(m-1)!}} | 0 \rangle
\end{equation}
where the director $\bf n$ is again a unit vector on $S^2$ given by 
$(\cos\phi \sin \theta, \sin \phi \sin \theta, \cos
\theta)$.
In this representation 
\begin{eqnarray}
\hat \rho &=& i \frac{\partial}{\partial \chi_k} \\ 
\hat {\bf S} &=& i {\bf n} \times \frac{\partial}{\partial {\bf n}} \\ 
\hat {\bf S}^2 &=& - \left[ \frac{1}{\sin \theta} \frac{\partial}{\partial
\theta} \left( \sin \theta \frac{\partial}{\partial \theta} \right) +
\frac{1}{\sin^2 \theta} \frac{\partial^2}{\partial \phi^2} \right] \\
\hat Q_{\alpha \beta} &=& N \left( {\bf n}_{\alpha} {\bf n}_{\beta} - 
\frac{1}{3}  \delta_{\alpha \beta} \right) 
\end{eqnarray}
The
Hamiltonian in Eq. 7 can be mapped to a Constrained Quantum Rotor Model 
(CQR), describing the dynamics of two unit vectors (${\bf n}, e^{i \chi}$) 
on a two-sphere and a unit circle:
\begin{equation}
\mathcal{H}_\text{CQR} = -t \sum_{\langle k l \rangle} {\bf n}_k \cdot 
{\bf n}_l
\cos (\chi_k - \chi_l)
+ \sum_k E_s \hat {\bf S}_k^2 + E_c \hat \rho_k^2 - \hat \rho_k \mu
\end{equation}
$t = N  \tilde t$. The CQR-model has been introduced to study spin-one 
bosons in a few previous works 
and we refer to those papers for detailed discussions.
\cite{Zhou02, Zhou01, Demler01}
For Mott states the effective Hamiltonian can be found as:
\begin{equation}
\mathcal{H}= E_s \sum_k {\bf S}_k^2 - J_{\text{ex}} \sum_{\langle k l 
\rangle}
({\bf n}_k \cdot {\bf n}_l)^2; \quad J_\text{ex} = \frac{t^2}{2 E_c},
\end{equation}
and we define $\eta=\frac{z J_{\text{ex}}}{E_s}$.

In general, we choose the on-site trial wave function to be:
\begin{equation}
\psi( {\bf n}_k ) = C_\sigma \exp \left[ \frac{\sigma}{2} ( {\bf n}_k 
\cdot {\bf n}_0)^2 \right].
\end{equation}
$C_\sigma$ is a normalization constant. When $\sigma \rightarrow 0 $ this
yields an isotropic state $Y_{00} ({\bf n}_k)$, which indicates a spin
singlet state. When $\sigma \rightarrow + \infty$, ${\bf n}_k$ is
localized on the two-sphere in the vicinity of ${\bf n}_0$, representing a
rod-like nematic state and when $\sigma \rightarrow - \infty$, ${\bf n}_k$
lies in a plane perpendicular to ${\bf n}_0$, corresponding to a disk-like
spin nematic state. Moreover
this wave function has the following property: $\psi(-{\bf n}_k)=\psi ({\bf
n}_k)$, as is
required for an even number of particles per site. \cite{Zhou01}

Choosing ${\bf n}_0 = e_z$ this gives:
\begin{equation}
\psi(\phi_k, \theta_k) = C_\sigma \exp \left[ \frac{\sigma}{2} \cos^2
\theta_k
\right].
\end{equation}

The expectation value of the Hamiltonian in this state is:
\begin{widetext}
\begin{eqnarray}
E_{\sigma} \!\! &=& \!\!
 E_s \! \left( \! - \frac{3}{4} -\frac{1}{2} \sigma + \frac{3
e^{\sigma} \sqrt{|\sigma|}}{2 \sqrt{\pi} {\rm Erfi} \sqrt{|\sigma}|} \!
\right)
 \!-\! z J_{\text{ex}} \!
 \left( \! \frac{ 12 e^{2 \sigma} \sigma -4 e^{\sigma}
 \sqrt{\pi} \sqrt{|\sigma|} (3 + 2 \sigma) {\rm Erfi} \sqrt{|\sigma|} +
 \pi( 3 + 4(\sigma+\sigma^2))
 {\rm Erfi}^2 \sqrt{|\sigma|}}{8 \pi \sigma^2 {\rm Erfi}^2 \sqrt{|\sigma|}}
 \! \right) \nonumber
\end{eqnarray}
\end{widetext}
in which ${\rm Erfi}[x]$ is the complex error function defined by ${\rm
Erf}[i x]/i$. In a series expansion for $\sigma \ll 1$, the result is:
\begin{eqnarray}
&& -\frac{z J_{\text{ex}}}{3}
   + \left(\frac{2}{15}E_s-\frac{8}{675} z J_{\text{ex}} \right) \sigma^2 \nonumber \\
&& + \left( \frac{4}{315} E_s - \frac{32}{14175} z J_{\text{ex}} \right) \sigma^3 \nonumber \\
&& + \left(-\frac{8}{4725} E_s +\frac{32}{165375} z J_{\text{ex}} \right) \sigma^4 + o(\sigma^{5})
\end{eqnarray}

\begin{figure}
\begin{center}
\psfrag{sigma}{$\sigma$}
\psfrag{EoverEs}{$E/E_s$}
\includegraphics[scale=.8]{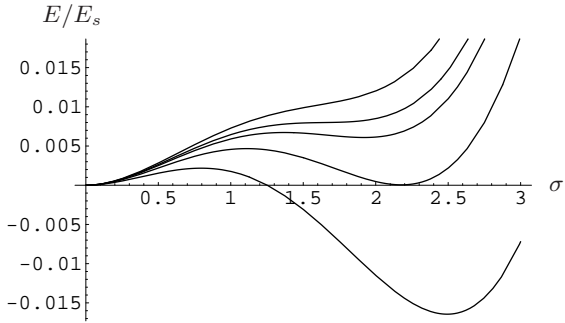}
\caption{Energy (in units of $E_s$) as a function of $\sigma$ for various
$\eta$ ($N =2k \gg 1$). From top
to bottom are curves for $\eta=9.9$, $9.96$, $10$, $10.0965$, $10.3$.}
\label{energysigma}
\end{center}
\end{figure}

The energy as a function of $\sigma$ at different $\eta$ is plotted in FIG.
\ref{energysigma}, which is qualitatively the same as FIG. \ref{energy}
for two particles per site.
When $\eta< 9.96$, the
energy  as a function of $\sigma$ has only one (global) minimum, which
corresponds to a spin singlet ground state.
When $\eta>9.96$, in addition to the global minimum, there appears a local
minimum at $\sigma > 0$.\
At $\eta = \eta_c = 10.0965$, these two minima become degenerate,
signifying a phase transition.
At $\eta>\eta_c$, the solution at $\sigma=0$ becomes a local minimum
indicating a metastable spin singlet state, whereas the global minimum at
$\sigma > 0$ corresponds to a nematic ground state.
As $\eta$ further increases, the solution at $\sigma=0$ becomes unstable 
and a local minimum occurs at $\sigma < 0$, while the global minimum 
remains at $\sigma>0$. Following discussions on Eqs. 32,33 we interpret 
the $\sigma<0$ solution as a metastable disk-like spin nematics. 

For the trial wave function in Eq. 44
the nematic order parameter can be calculated as:
\begin{eqnarray}
\tilde Q &=& \frac{ \langle \sigma | \hat Q_{\alpha \beta} | \sigma
\rangle}{\langle
\infty | \hat Q_{\alpha \beta} | \infty \rangle} \nonumber \\
&=& -\frac{1}{2} - \frac{3}{4 \sigma} +\frac{3 e^{\sigma}}{2 \sqrt{\pi
|\sigma| }
{\rm Erfi} \sqrt{|\sigma|}}
\end{eqnarray}
When $\tilde Q$ is small, we obtain an expression of energy in terms
of $\tilde Q$:

\begin{equation}
E_{\tilde Q}  = -\frac{z J_{\text{ex}}}{3} + \left(
\frac{15}{2} E_s - \frac{2}{3} z J_{\rm ex} \right) \tilde Q^2 -
\frac{75}{14}
E_s \tilde Q^3  + \frac{1275}{98} E_s \tilde Q^4
\end{equation}
The jump in $\tilde Q$ at the phase transition is equal to $0.323$.

\begin{figure}
\begin{center}
\psfrag{tilde_Q}{$\tilde Q$}
\psfrag{eta}{$\eta$}
\includegraphics[scale=.8]{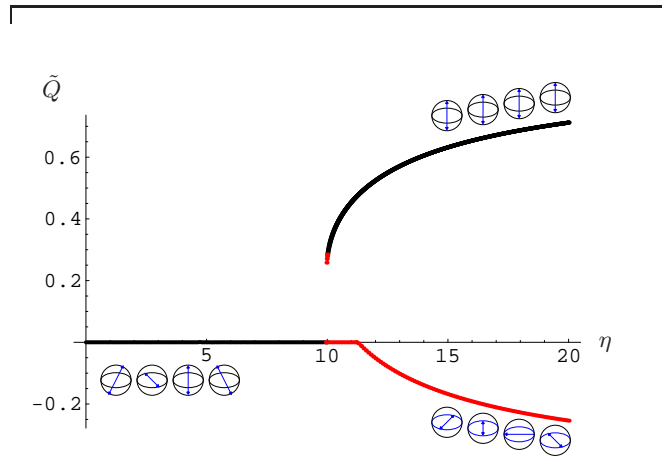}
\caption{(Color online) Nematic order parameter as a function of $\eta$ for $N=2k (\gg
1)$. Along
the black lines are ground states; along the red (light) lines are metastable
states.
(See also the caption of FIG. 2.)}
\label{qetalarge}
\end{center}
\end{figure}

The evolution of ground state wave functions and results on quantum phase
transitions are summarized in figure \ref{qetalarge}, where the nematic
order parameter is plotted as a function of $\eta$.
As stated before, these results are only valid in high dimensional lattices,
where fluctuations in ordered states are small. For detailed 
calculations of fluctuations we refer to appendix \ref{fluctuations}.

\subsection{Large $N$ limit: An odd number of particles per site}
At last, we also present results for an odd number of atoms
per site. The main difference between this case and the case for an
even number of particles per site is that at zero hopping limit
in the former case there is always an unpaired atom at each site.
Consequently in the mean field approximation, we only find nematic Mott
insulating phases. As in the case for even numbers of particles per site,
we expect this approximation to be valid in high
dimensional lattices but fail in low dimensions, especially
in one-dimensional
lattices where long wave length fluctuations are substantial.
Here we restrict ourselves to high dimensional lattices only.

For large $N$ a trial wave function which interpolates between spin
singlet states
(dimerized) and nematic states can be introduced as:
\begin{eqnarray}
\Psi_{\text{odd}} (\{ {\bf n}_k \}) &=& \prod_{\langle k l \rangle_{\text
p}}
C(O, \sigma)
\left[ O ({\bf n}_k \cdot {\bf n}_0) ( {\bf n}_{l} \cdot {\bf n}_0)+
( {\bf n}_k \cdot {\bf n}_{l}) \right]  \nonumber \\
& &  \times \exp \lbrack \sigma (
({\bf n}_k \cdot {\bf n}_0)^2+({\bf n}_{l} \cdot {\bf n}_0)^2) \rbrack.
\end{eqnarray}
$\langle  k l \rangle_{\text p}$ denotes that the summation should be
taken over parallely ordered pairs of nearest neighbors $k$ and $l$
covering the lattice.  $C(O, \sigma)$ is a normalization constant.
The solution with $O=0, \sigma=0$ corresponds to a dimerized valence bond
crystal
state;
and solutions with $O \neq 0 $, or $\sigma\neq 0$ represent nematic
states.

\begin{figure}
\begin{center}
\psfrag{eta}{$\eta$}
\psfrag{O}{$O$}
\includegraphics[scale=.8]{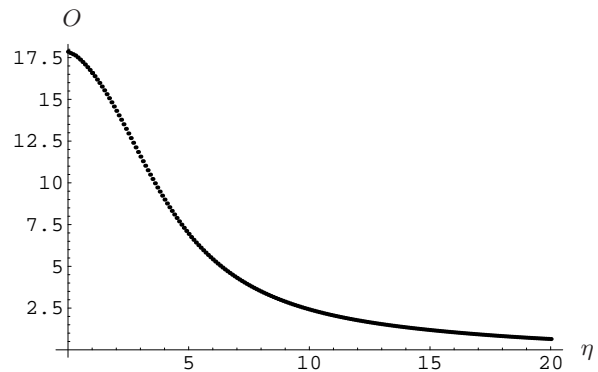}
\caption{The value of $O$ as a function of $\eta$ ($N=2k+1 \gg 1$).}
\label{aodd}
\end{center}
\end{figure}

\begin{figure}
\begin{center}
\psfrag{eta}{$\eta$}
\psfrag{sigma}{$\sigma$}
\includegraphics[scale=.8]{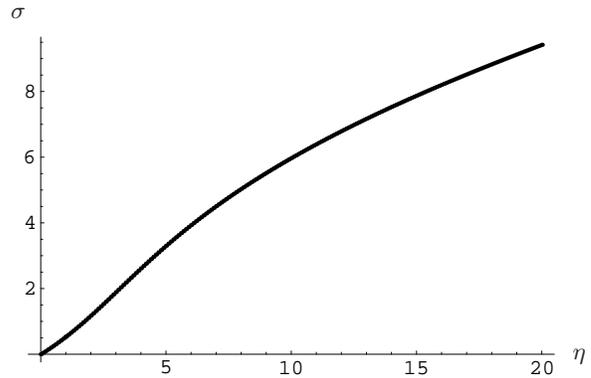}
\caption{The value of $\sigma$ as a function of $\eta$ ($N=2k+1 \gg 1$).}
\label{sigmaodd}
\end{center}
\end{figure}

It is straightforward, but tedious to compute the energy of these states.
Minimizing it with respect to various values of $\eta$ for $d=3$ gives the
results shown in FIG. \ref{aodd} and \ref{sigmaodd}.
No phase transitions are found in the mean field
approximation;
and ground states break both rotational and translational symmetries.
\cite{Yip03}

At very small $\eta$, the on-site Hilbert space
is truncated into the one for a spin-one particle. \cite{Zhou02} The
reduced
Hamiltonian
in the truncated space
is a Bilinear-Biquadratic model for spin-1 lattices

\begin{equation}
{\cal H}_{\text{b.b.}}=J\sum_{ \langle kl \rangle} [ \cos\theta {\bf S}_k\cdot
{\bf S}_l +\sin\theta ({\bf S}_k \cdot {\bf S}_l)^2 ], {\bf S}_k^2=2;
\end{equation}
$\theta$ in general varies between $-3\pi/4$ and $-\pi/2$.
We therefor expect ground states at small $\eta$ limit should still
exhibit nematic order (i.e. $O\neq 0$).

It is worth emphasizing that conclusions about small $\eta$ limit arrived
here
are only valid in high dimensional bipartite lattices.
In low dimensional lattices, states of correlated atoms in this limit were
discussed recently and
ground states could be rotationally invariant dimerized-valence-bond
crystals. \cite{Zhou02}

\section{Conclusions}
We have studied the microscopic wave functions of spin nematic and spin 
singlet Mott states.
Both disk-like and rod-like spin nematic states were investigated. 
We also have analyzed quantum phase transitions between spin singlet
Mott insulating states and nematic Mott insulating states. We show that 
in the mean field
approximation, the phase transitions are weakly first order ones.
Thus, we expect that fluctuations play a very important role in
these transitions and the full theory on quantum phase transitions
remains to be discovered.

On the other hand, we have estimated fluctuations in different regimes of
the parameter space. We found that  fluctuations are indeed small away 
from the critical point, at either small hopping or large hopping limit 
for an even number of particles per site.
At the small hopping limit, fluctuations are proportional to $\eta$,
while at the large hopping limit they can be estimated to be proportional 
to $\frac{1}{\sqrt{\eta}}$ (see Appendix B).

For an odd number of particles per site, fluctuations are small only
at large hopping limit and are significant at small hopping limit.
The later fact implies a large degeneracy of Mott states at zero hopping
limit which was emphasized in the discussions on low dimensional Mott
states. The physics in this limit remains to be fully understood.

In the context of antiferromagnets, spin nematic states have also been 
proposed. \cite{Andreev84, Chubukov90, Chandra90} Collective
excitations in atomic nematic states
should be similar to those studied in previous works; we present some
brief discussions on this subject in Appendix B and refer to 
\cite{Andreev84, Chubukov90, Chandra90} for details.

\section{Acknowledgment}
This work is supported by the foundation FOM, in the Netherlands under
contract 00CCSPP10, 02SIC25 and NWO-MK projectruimte 00PR1929;
MS is also partially supported by a grant from Utrecht University.
\\
\\
{\bf Note added}: At a later stage of this work, we received a copy of
the manuscript by A. Imambekov, M. Lukin and E. Demler where similar
results have been obtained.

\appendix
\section{An alternative description}
Alternatively, one can also carry out the calculations in section III,
using the following operator:
\begin{eqnarray}
\hat Q_{2, \alpha \alpha'} &=& \hat {\bf S}^\alpha \hat {\bf S}^{\alpha'}
-\frac{1}{3} \delta_{\alpha \alpha'} \hat {\bf S}^\gamma \hat {\bf
S}^{\gamma}
\nonumber \\ &=&
- \epsilon^{\alpha \beta \gamma} \epsilon^{\alpha' \beta' \gamma'}
\psi_\beta^\dagger \psi_{\beta'}^\dagger \psi_{\gamma} \psi_{\gamma'} +
\delta_{\alpha \alpha'} \psi_\eta^\dagger \psi_\eta
\nonumber \\ &&- \psi_{\alpha'}^\dagger
\psi_\alpha -\frac{1}{3} \delta_{\alpha \alpha'} \hat {\bf
S}_{\text{tot}}^2
\end{eqnarray}
Defining the more conventional order parameter \cite{Andreev84}
\begin{equation}
\tilde Q_2 = \frac{\langle \hat  Q_{2, \alpha \alpha'} \rangle}{
\langle \hat Q_{2, \alpha \alpha'}\rangle_{\text{ref}}},
\end{equation}
we obtain the following results for the trial wave function in Eq. 28:
\begin{eqnarray*}
\tilde Q_2 &=& \sqrt{\frac{3}{2}} \sin^2  \theta \\
E &=& 6 E_s \sqrt{\frac{2}{3}} \tilde Q_2  \\
&&-  \frac{2}{3} z \tilde J_{\text{ex}}
\left( 2 \sqrt{2} \sqrt{ \sqrt{\frac{2}{3}}(1- \sqrt{\frac{2}{3}} \tilde
Q_2 )
\tilde Q_2} + \sqrt{\frac{2}{3}} \tilde Q_2 \right)^2 \\
&=& \left( 2 \sqrt{6} E_s - \frac{16}{3}\sqrt{\frac{2}{3}} z 
\tilde J_{\text{ex}} \right) \tilde Q_2 \\
&& - \frac{16 \sqrt[4]{2}}{\sqrt[4]{3^7}} z J_{\text{ex}} \tilde Q_2^{3/2} 
+ \frac{28}{9} z \tilde J_{\text{ex}} \tilde Q_2^2 + O(\tilde Q_2^{5/2})
\end{eqnarray*}
which lead to the same conclusions as in section III. However, 
in terms of the order parameter defined in Eq. A2, the rod-like and 
disk-like 
structures shown in FIG.2 and FIG.4 are less obvious.

In the case of a large number of particles per site, the
order parameter introduced here has the same expectation value as 
the operator in Eq. 40. 

\section{spin fluctuations in Mott states} \label{fluctuations}
Nonlinear dynamics and spin fluctuations in condensates of spin-one bosons
were discussed in a previous work. \cite{Zhou01}
Here we carry out a similar discussion for Mott states.
Following the Hamiltonian
\begin{equation}
\mathcal{H} = E_s \sum_k \hat {\bf S}_k^2 - J_{\text{ex}} \sum_{\langle k l
\rangle} ({\bf n}_k \cdot {\bf n}_l)^2
\end{equation}
we derive the following equation of motion for the director ${\bf n}_k$:

\begin{equation}
\frac{d {\bf n}_k}{d t} = 2E_s  \hat {\bf S}_k \times {\bf n}_k 
\end{equation}

{\bf Fluctuations when $\eta$ is small} 

For $\eta={z J_{\text{ex}}}/{E_s}=0$ and an even number of particles 
per site, the ground state is the product state:
\begin{equation}
\Psi_{\eta=0} = \prod_k Y_{00} ( {\bf n}_k)
\end{equation}
When $0<\eta\ll 1$,  
the ground state wave function can also be obtained by a perturbation 
theory; the leading term is

\begin{equation}
\Psi_0^{(1)} = \sum_{l \neq 0,m} \frac{ \langle \Psi_{lm}^{(0)} | 
-J_{\text{ex}} 
\sum_{ \langle kl \rangle} ({\bf n}_k\cdot {\bf n}_l)^2
| \Psi_{00}^{(0)} \rangle}{E_{00}^{(0)} - E_{lm}^{(0)}}
\end{equation}
In our case $\Psi_{lm}^{(0)}=Y_{lm}$ with $l$ even, 
and $E_{l}^{(0)}=l(l+1)E_s$. 
A direct calculation yields

\begin{equation}
\Psi_0^{(1)} = \frac{\eta}{45 z} \sum_{\langle i j \rangle} \sum_{m=-2}^2
Y_{2 m} ({\bf n}_i) Y_{2, -m} ({\bf n}_j) \prod_{k\neq i, j} Y_{00} ({\bf 
n}_k).
\label{1st}
\end{equation}

Taking into account $\langle Y_{00} | \hat Q_{\alpha \beta} | 
Y_{00}\rangle=0$, we find desired results in this limit,
\begin{equation}
\langle  \hat Q_{k, \alpha \beta} \rangle=0.
\end{equation}

To characterize fluctuations, we study the following correlation function
$\langle  \hat Q_{k, \alpha \alpha} \hat Q_{k', \alpha \alpha} \rangle$. 
Calculations of this correlation function in the state given in 
Eq. \ref{1st} yield
\begin{multline}
\langle  \hat Q_{k, \alpha \alpha} \hat Q_{k', \alpha \alpha} \rangle 
=  
\frac{2 \eta}{45}  \delta({kk',\langle kl \rangle}) \times \\ 
\sum_{m=-2}^{2} 
\left( \langle Y_{2m} | \hat Q_{\alpha \alpha} | Y_{00} \rangle 
\langle Y_{2,- m} | \hat Q_{\alpha \alpha} | Y_{00} \rangle 
+ \text{h.c.}\right). \nonumber
\end{multline}
$\delta({kk', \langle kl \rangle})$ is unity if $k'$ and k sites are two 
neighboring sites 
as $\langle kl \rangle$ and otherwise is zero.
The last expression can be calculated explicitly, 
\begin{equation}
\left( \langle Y_{2m} | \hat Q_{\alpha \alpha} | Y_{00} \rangle 
\langle Y_{2,- m} | \hat Q_{\alpha \alpha} | Y_{00} \rangle 
+ \text{h.c.} 
\right)
= \frac{8}{45} 
\end{equation}
Clearly at small $\eta$, fluctuations are small.

{\bf Fluctuations when $\eta$ is large}

Again we consider the case for an even number of particles per site. In 
the
limit of $\eta \rightarrow \infty$, all directors ${\bf n}_k$ 
point in the direction of ${\bf e}_z$. For a finite but large 
$\eta$ we introduce

\begin{equation}
{\bf n}_k={\bf e_z}\sqrt{1-C^2_{kx}-C^2_{ky}}
+C_{kx} {\bf e}_{x} +C_{ky} {\bf e}_{y}
\end{equation}
where $C_{k\alpha}$, $\alpha=x,y$ are much less than unity.

Following discussions in section IIIB, we obtain the following 
commutators,
\begin{equation}
\lbrack \hat {\bf S}_{ky},  C_{k'x} \rbrack 
\approx i \delta_{k,k'},  
\lbrack \hat {\bf S}_{kx},C_{k'y} \rbrack 
\approx -i\delta_{k,k'}   
\end{equation}
which define two sets of harmonic oscillators. 
Introducing 

\begin{equation}
\hat 
\Pi_y= \hat {\bf S}_x, \hat \Pi_x=-\hat {\bf S}_y
\end{equation}
the effective Hamiltonian becomes
\begin{equation}
\mathcal{H} = \sum_{\alpha = x, y} 
[ E_s \sum_k \hat \Pi_{k,\alpha}^2 +  J_{\text{ex}} \sum_{\langle k l 
\rangle} ({C}_{k, \alpha}-C_{l, \alpha})^2].
\end{equation}
And 
\begin{equation}
[\hat{\Pi}_{k\alpha}, C_{k'\beta}]=i
\delta_{k,k'}\delta_{\alpha\beta}.
\end{equation}

To obtain eigenmodes, we perform a Fourier transformation (setting the
lattice 
spacing to be unity),

\begin{equation}
\hat{\Pi}_{k, \alpha}=\frac{1}{\sqrt{V_T}}\sum_{\bf q} \hat \Pi_{{\bf q}, 
\alpha} e^{i 
\pi k \cdot {\bf q}}, 
{C}_{k, \alpha}
 =\frac{1}{\sqrt{V_T}} \sum_{\bf q} C_{{\bf q}, \alpha}  e^{i \pi 
k\cdot {\bf q}}, 
\end{equation}
where $V_T$ is the total number of lattice sites.
This leads to the following
Hamiltonian
\begin{equation}
\mathcal{H} = \sum_{{\bf q}, \alpha}[ E_s \hat \Pi_{\bf {q}, \alpha}^2 + z 
J_{\text{ex}} \sin^2 \frac{|{\bf q}|\pi}{2} C_{{\bf q}, \alpha}^2]. 
\end{equation}   

Following a standard calculation,  fluctuations in this limit are:

\begin{equation}
\langle \sum_{\alpha} C^2_{k,\alpha} \rangle =
\frac{1}{V_T} \langle \sum_{{\bf q},\alpha}|C_{{\bf q},\alpha}|^2 \rangle
=\frac{2}{\sqrt{\eta}}\frac{1}{V_T}
\sum_{|{\bf q}| < q_c}\frac{1}{\sin |{\bf q}|\pi/2}.
\label{flu}
\end{equation}
The momentum cut-off $q_c$ in general depends on the short distance
behavior of our model and for simplicity we set it as one.
In high dimensional lattices, the sum in Eq. \ref{flu} is convergent;
and we see the fluctuations are also small at the large $\eta$ limit.

\end{document}